\documentclass[%
 aip,
 amsmath,amssymb,
 reprint,%
groupedaddress
]{revtex4-1}

\usepackage{graphicx}% Include figure files
\usepackage{dcolumn}% Align table columns on decimal point
\usepackage{bm}% bold math
\usepackage[T1]{fontenc}
\usepackage[english]{babel}
\usepackage{mathptmx}
\usepackage{xcolor}

\newcommand{\be}{\begin{equation}}
\newcommand{\ee}{\end{equation}}
\newcommand{\bea}{\begin{eqnarray}}
\newcommand{\eea}{\end{eqnarray}}

\begin{document}

%  \preprint{AIP/123-QED}

\title[]{Comment on "Flexibility of short DNA helices with finite-length
effect: From base pairs to tens of base pairs"}
% Force line breaks with \\

\author{Midas Segers}
\author{Enrico Skoruppa}
\author{Jan A. Stevens}
\author{Merijn Vangilbergen}
\author{Aderik Voorspoels}
\author{Enrico Carlon}
\email{enrico.carlon@kuleuven.be}
\affiliation{Soft Matter and Biophysics, KU Leuven, Celestijnenlaan 200D, 3001 Leuven, Belgium}

\date{\today}% It is always \today, today,

\begin{abstract}
While analyzing the persistence length of DNA atomistic simulations
Wu {\sl et al.} [J. Chem. Phys. {\bf 142}, 125103 (2015)] introduced
an empirical formula to account for the observed length-dependence. In
particular they found that the persistence length increases with the
distance.  Here, we derive the formula by Wu {\sl et al.} using a
non-local twistable wormlike chain which introduces couplings between
distal sites.  Finally, we show that the same formula can account for
the length-scale dependence of the torsional persistence length and is,
in fact, applicable to any kind of polymer model with non-local couplings.
\end{abstract}

\maketitle

%%%%%%%%%%%%%%%%%%%%%%%%%%%%%%%%%%%%%%%%%%%%%%%%%%%%%%%%%%%%%%%

In Ref.~\onlinecite{wu15} Wu {\sl et al.} analyzed the bending flexibility
of short DNA segments (of up to 50 bps) by molecular dynamics and Monte
Carlo simulations. They found that DNA is softer at short lengths,
which they accounted for with the empirical formula
\begin{equation}
l_B (m) = l_B^\infty - \frac{A}{B+m},
\label{lB_empirical}
\end{equation}
where $l_B (m)$ is the persistence length of a sequence of $m$ base
pair-steps, $l_B^\infty$ the asymptotic persistence length and $A$, $B$
some fitting parameters. In this comment we show that the above formula
is consistent with the asymptotic expansion of the non-local Twistable
Wormlike chain (nlTWLC) \cite{skor21} which gives
\begin{equation}
l_B (m) = l_B^\infty \left( 1 - \frac{B}{B+m} \right) + 
%  {O}\left( e^{-m/\lambda} \right),
\ldots
\label{lB_exact}
\end{equation}
%  with the last term indicating exponentially small corrections.  
with the dots indicating higher order corrections. Such expansion is
valid both for (1) the local persistence length of a long DNA molecule
measured from the tangent-tangent correlation of two base pairs
separated by $m$ steps and for (2) the persistence length obtained
from the correlation of the end-point tangent vectors for a DNA
molecule of finite length $m$. The latter is the quantity analyzed by
Wu et al.\cite{wu15}. %We will also show 
Furthermore, we will show,
that an expansion of the type
\eqref{lB_exact} is also valid for the torsional persistence length.
Note, that expression \eqref{lB_exact} contains one parameter less than
\eqref{lB_empirical} implying that $A = l_B^\infty B$. This relation
holds approximately for the values reported by Wu et al. \cite{wu15}:
$l_B^\infty =50$~nm, $A=450$~nm and $B=10$.

{\sl Non-local couplings --} The nlTWLC describes DNA configurations
using the three rotational densities tilt, roll (the two bending modes),
and twist indicated with $\tau_n$, $\rho_n$, and $\Omega_n+\omega_0$
respectively, with $n=1,2,\ldots N$ enumerating the base pair-steps
along the chain.  The twist is subdivided into a constant intrinsic twist
density $\omega_0 =1.75~\text{nm}^{-1}$, corresponding to one turn of the
helix every $10.5$ base pairs, and a fluctuating excess twist $\Omega_n$.
While in the ordinary TWLC couplings are strictly local and the energy
is quadratic in $\tau_n$, $\rho_n$, and $\Omega_n$, the nlTWLC introduces
couplings between distal sites, such as for instance $\tau_n \tau_{n+m}$
with $m>0$.  
Non-local couplings have been observed in several all-atom simulations
\cite{lank03,noy12} and they typically extend to three flanking dinucleotide
steps\cite{skor21}.
%  In principle, such couplings can extend to an arbitrary range, however
%  simulations indicate that in DNA these are limited to a few flanking
%  dinucleotide steps \cite{skor21}.
%  The nlTWLC can be conveniently
%  described using a momentum space representation, introducing the discrete
%  Fourier transforms of the deformation parameters $\widetilde\tau_q$,
%  $\widetilde\rho_q$ and $\widetilde\Omega_q$ with $ -(N-1)/2 \leq q \leq
%  (N-1)/2$, for $N$ odd. The energy (in units of $k_BT$) of this model
%  then becomes \cite{skor21}
%  \begin{equation}
%  \beta E = \frac{a}{2N} \sum_q \Delta_q^\dagger \widetilde{M}_q \Delta_q,
%  \end{equation}
%  where $a=0.34$~nm is the base pair distance, $\Delta_q^T =
%  (\widetilde\tau_q, \widetilde\rho_q, \widetilde\Omega_q)$ the momentum
%  space strain-field vectors and $\widetilde{M}_q$ the associated $3 \times
%  3$ stiffness matrices.  The precise form of these matrices encodes the
%  range of distal site couplings, see Ref.~\onlinecite{skor21}.

{\sl Persistence lengths --} The nlTWLC predicts similar length scale
dependence of both the bending and the torsional persistence length,
hence we discuss both. In the limit of an infinite long chain
$N \to \infty$ the correlation functions between two sites separated by
$m$ base pairs decay with $m$-dependent persistence lengths given by
the following expressions\cite{skor21}:
  \begin{equation}
  \frac{1}{l_\chi} = \frac{1}{m\pi}  \int_{-\pi/2}^{+\pi/2}
  \frac{\sin^2 (my)}{\sin^2 y} \, f_\chi (y) \, dy,
  \label{DNA:lB2}
  \end{equation}
with $\chi=\{B,T\}$ labeling either the bending or the torsional
persistence length. The variable $y=\pi q/N$ is the rescaled momentum and
$f_\chi (y)$ are functions of the bending fields (tilt and roll) and 
the excess twist field for the $l_B$ and $l_T$ respectively. The full
expressions are reported in Ref.~\onlinecite{skor21}. We also note, that 
while the former expression contains certain approximations, the latter 
is exact.
%the functions $f_\chi (y)$ are different in the two cases\cite{skor21}:
%$f_\text{B}(y)$ can be expressed as a combination of thermal averages of
%bending degrees of freedom tilt and toll and $f_\text{T}(y)$ of the excess
%twist.  
%The expressions for $l_\text{T}$ and $l_\text{B}$ are reported in
%Ref.~\onlinecite{skor21}. 
% I WOULD PRESERVE THE ORDER OF L_T AND L_B TO AVOID CONFUSION
%We also note that while the former is exact,
%the latter contains some approximations.  
The functions $f_\chi(y)$
are symmetric in $y$, so for small $y$ the following expansions hold
\begin{equation}
f_\chi(y) \approx f_\chi(0) + \frac{1}{2} f_\chi''(0)\, y^2 + 
O\left( y^4 \right).
\label{fy_exp}
\end{equation}
To proceed further it is convenient to split $f_\chi(y)$ in
Eq.~\eqref{DNA:lB2} into two terms, a constant $f_\chi(0)$ and the
increment $f_\chi(y)-f_\chi(0)$ and integrate these two terms separately.
That way, one obtains (to simplify the formula we omitted the integration
boundaries)
\begin{eqnarray}
\frac{1}{l_\chi} &=& 
\frac{f_\chi(0)}{m\pi} \int \frac{\sin^2 my}{\sin^2 y} dy +
\int \frac{1-\cos (2 m y)}{2 m \pi}  \frac{f_\chi(y)-f_\chi(0)}{\sin^2 y} dy
\nonumber \\
&=& f_\chi(0) \left( 1 + \frac{B_\chi}{m} \right) + 
O\left( e^{-m/\lambda} \right),
\label{lB:expansion}
\end{eqnarray}
where we have used 
\begin{eqnarray}
\int_{-\pi/2}^{\pi/2} \frac{\sin^2 my}{\sin^2 y} \, dy &=& m \pi,
\label{app:int_sinmsin}
\end{eqnarray}
and the definition
\begin{eqnarray}
 B_\chi &\equiv& \frac{1}{2 \pi f_\chi(0)} \int_{-\pi/2}^{+\pi/2}
\frac{f_\chi(y)-f_\chi(0)}{\sin^2 y} dy.
\label{defB}
\end{eqnarray}
The oscillatory term $\cos (2my)$ in \eqref{lB:expansion} yields an
exponentially small factor proportional to $\exp(-m/\lambda)$, where the
characteristic decay length $\lambda$ can be obtained from the poles of
the integrand \cite{skor21}. We note that the integral in \eqref{defB} is
convergent, in particular \eqref{fy_exp} implies that the integrand does
not diverge as $y \to 0$. Finally, inverting Eq.~\eqref{lB:expansion}
one gets \eqref{lB_exact} with $l_B^\infty = 1/f_B(0)$ and $B=B_B$. 
%  One can perform a similar analysis with a real space calculation, which
%  shows that the term proportional to 1/m in Eq.(8) comes from a boundary
%  contribution for a space correlator of fields in a domain of length m.

While so far we have considered the limit $N \to \infty$ we can
extend the analysis for a sequence of finite length. For 
%simplicity 
brevity
we
limit the discussion to the bending persistence length. For 
finite $N$ this is given by\cite{skor21}
\begin{eqnarray}
\frac{1}{l_\text{B}} = \frac{a}{2m} \left\langle
\left(\sum_{n=0}^{m-1} \tau^*_n \right)^2 +
\left(\sum_{n=0}^{m-1} \rho^*_n \right)^2
\right\rangle
\label{lB:finite}
\end{eqnarray}
where $\tau^*_m$ and $\rho^*_m$ are related to the original variables
tilt and roll by a linear transformation\cite{skor21}. 
%In particular, we
%are interested, as in Ref.~\onlinecite{wu15}, 
In particular, as in Ref.~\onlinecite{wu15}, we
are interested,
in the case $m=N$, where
$l_\text{B}$ is obtained from the tangent-tangent correlation function
between the two ends of the chain. On general grounds the averages in
\eqref{lB:finite} will produce a bulk type term extensive in $m$ and
boundary corrections %as
$\langle X \rangle = m \Gamma_1 + \Gamma_2 +
\ldots$, with $X$ indicating the terms averaged in \eqref{lB:finite}. Such
expression produces again the expansion \eqref{lB_exact}.

%  We note that in a local
%  model off-site correlators vanish $\langle \tau^*_n \tau^*_{n+k} \rangle
%  = 0$  for $k > 0$ and the sums in \eqref{lB:finite} are proportional
%  to $m$ so that the $m$-dependence cancels for $l_\text{B}$.  
%  Off-site interactions will give non-vanishing correlations. On general ground
%  for large $m$
%  \begin{eqnarray}
%  \left\langle
%  \left(\sum_{n=0}^{m-1} \tau^*_n \right)^2 +
%  \left(\sum_{n=0}^{m-1} \rho^*_n \right)^2
%  \right\rangle = m \Gamma_1 + \Gamma_2 + \ldots
%  \label{lB:finite2}
%  \end{eqnarray}

%%%%%%%%%%%%%%%%%%%%%%%%%%%%%%%%%%%%%%%%%%%%%%%%%%%%%%%%%%%%%%%%%%%%%%%%
\begin{figure}[t]
\includegraphics[width=0.45\textwidth,angle=0]{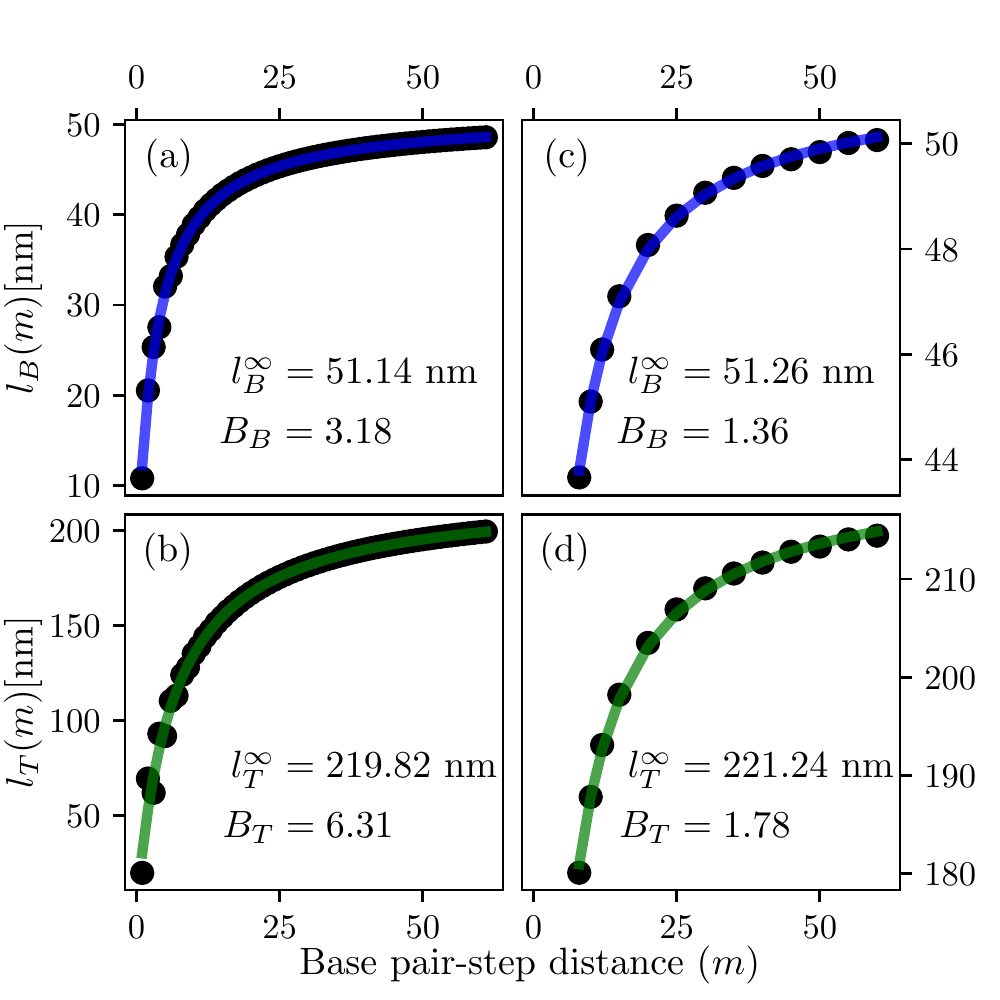} 
\caption{Bending and torsional persistence lengths deduced from Monte
Carlo simulations of the nlTWLC (black dots).  The left-hand panels show
$m$-dependent bending (a) and torsional (b) persistence lengths for a
chain of length $N=200$, while the right-hand panels show end-to-end
bending (c) and torsional (d) persistence length ($m=N$) for various
chain-lengths.  Non-local energetic couplings are included up to third
closest neighbors and are chosen to induce significant length-scale
dependence, parameters are given in Table~\ref{app:tabel_MC}. Fits of
\eqref{lB_exact} to the respective correlation lengths are shown as blue
and green lines.}
\label{FIG}
\end{figure}
%%%%%%%%%%%%%%%%%%%%%%%%%%%%%%%%%%%%%%%%%%%%%%%%%%%%%%%%%%%%%%%%%%%%%%%%

{\sl Monte Carlo simulations--}
Figure~\ref{FIG} shows an example of both torsional and bending
persistence lengths obtained from Monte Carlo simulations of the
nlTWLC compared to fits with Eq.~\eqref{lB_exact} (solid lines).  
%  Note, that the exponentially small correction contains an oscillatory
%  factor $(-1)^m$ and vanishes rapidly with increasing $m$.  This behaviour
%  is reflected by the deviations between the Monte Carlo data and
%  Eq.~\eqref{lB_exact} for small $m$.
Left plots are for $N=200$ and finite $m$, while the right plot for
sequences of finite length $m=N$.  As expected the asymptotic persistence
lengths are the same in these two cases, as both converge towards the
same $l^\infty_\text{B,T}$ for large $m$. However, the $B_\chi$ are
determined by the boundary contributions of short segments (small $m$).
These contributions are manifestly different when comparing a partial
segments located within a larger chain ($m \ll N$) and segments spanning
the entire chain ($m=N$), as the latter case lacks any couplings spanning
beyond the considered range.

%strongly differ, because the $B_\Chi$ are 
%determined by the boundary terms of short segments. 
%
% However, $B_$  the values of $B$ are different 

{\sl Static contributions --}
Thus far we have solely considered models in which the persistence length
only depends on the thermal bending fluctuations of the molecule. However,
it is well established, that the ground-state of a DNA molecule is not
straight, but exhibits significant static bending. These bending
components will further expedite the decorrelation of tangents. It
has been shown\cite{trif87}, that the bending peristence length can
be decomposed into the dynamic component $l_D$ determined by thermal
fluctuations and the static component $l_S$ as
\begin{equation}
\frac{1}{l_\text{B}} = 
\frac{1}{l_D} + \frac{1}{l_S}.
\end{equation}
When considering a sufficiently large chain, one can define the average
$m$-dependent static peristence length by considering a sequence ensemble.
We assume that in this ensemble averaging produces uncorrelated static
bends. Following this assumption $l_S$ is taken $m$-independent,
while for the dynamic component (which is of thermal origin)
Eq.~\eqref{lB:expansion} still holds. Hence we find
\begin{eqnarray}
\frac{1}{l_B} &=& \frac{1}{l_D^\infty}\left(1+\frac{B_D}{m}\right) + 
\frac{1}{l_S},
\end{eqnarray}
which ultimately gives Eq.~\eqref{lB_exact} with
\begin{eqnarray}
\frac{1}{l_B^\infty} = \frac{1}{l_D^\infty} + \frac{1}{l_S}  
\quad\quad&\text{and}&\quad\quad
B = \frac{l_B^\infty}{l_D^\infty} \, B_D.
\end{eqnarray}

%%%%%%%%%%%%%%%%%%%%%%%%%%%%%%%%%%%%%%%%%%%%%%%%%%%%%%%%%%%%%%%%%
\begin{table}[t]
\caption{Real-space couplings used in the Monte Carlo simulation shown
in Fig.~\ref{FIG}. $X_k$ indicates the couplings between sites displaced
by $k$ base pair-steps. 
%  The bending modes were chosen to be isotropic,
%  i.e. $A_1 = A_2$ for all $k$ and no local couplings between different
%  rotational modes, such as twist-bend coupling ($G$) were included.
We used an isotropic model with bending stiffness $A$ and torsional stiffness
$C$.
The asymptotic bending- and twist-stiffness according to Eq.~\eqref{lB_exact}
are indicated by $X^\infty$. For the intrinsic twist density and
discretization length $\omega_0 = 1.75$ nm$^{-1}$ and $a=0.34$ nm were
used for respectively. Since the bending modes are expressed as rotational
densities the the couplings are expressed as length in units of nm.}
\begin{ruledtabular}
\begin{tabular}{l|ccccc}
& $X_0$ & $X_1$ & $X_2$ & $X_3$ & $X^\infty$ \quad\qquad \\
\hline
$A$ & $25$ & $14$ & $5$ & $2$ & $51$  \quad \qquad \\
$C$ & $40$ & $25$ & $8$ & $2$ & $110$ \quad \qquad \\
\end{tabular}
\end{ruledtabular}
\label{app:tabel_MC}
\end{table}
%%%%%%%%%%%%%%%%%%%%%%%%%%%%%%%%%%%%%%%%%%%%%%%%%%%%%%%%%%%%%%%%%

%  \bibliography{references.bib}


\end{document}